\documentclass[aps,pra,showpacs,superscriptaddress,preprint]{revtex4}
\usepackage{amssymb}
\usepackage{amsmath}
\usepackage{graphicx}

\catcode`ð=\active
 \defð{\u{g}}
 \catcode`Ð=\active
 \defÐ{\u{G}}
 \catcode`Ý=\active
\defÝ{\. I}
 \catcode`ö=\active
\defö{\"{o}}
 \catcode`Ö=\active
 \defÖ{\"O}
 \catcode`ü=\active
 \defü{\"{u}}
 \catcode`Ü=\active
 \defÜ{\"{U}}
 \catcode`Þ=\active
\defÞ{\c{S}}
 \catcode`þ=\active
 \defþ{\c{s}}
 \catcode`ý=\active
 \defý{{\i}}
 \catcode`ç=\active
\defç{\d{c}}
 \catcode`Ç=\active
\defÇ{\d{C}}

\input{tcilatex}

\begin{document}

\title{Relativistic and nonrelativistic bound states of the isotonic
oscillator by Nikiforov-Uvarov method }
\author{Sameer M. Ikhdair}
\email[E-mail: ]{sikhdair@neu.edu.tr}
\affiliation{Department of Physics, Near East University, Nicosia, North Cyprus, Turkey}
\author{Ramazan Sever}
\email[E-mail: ]{sever@metu.edu.tr}
\affiliation{Department of Physics, Middle East Technical University, 06800, Ankara,Turkey}
\date{%
\today%
}

\begin{abstract}
A nonpolynomial one-dimensional quantum potential in the form of an isotonic
oscillator (harmonic oscillator with a centripetal barrier) is studied. We
provide the non-relativistic bound state energy spectrum $E_{n}$ and the
wave functions $\psi _{n}(x)$ in terms of the associated Laguerre
polynomials in the framework of the Nikiforov-Uvarov method. Under the spin
and pseudospin symmetric limits, the analytic eigenvalues and the
corresponding two-component upper- and lower-spinors of the Dirac particle
are obtained, in closed form.

\end{abstract}

\maketitle

\newpage

\section{Introduction}

The exact solutions of the nonrelativistic and relativistic wave equations
are only possible in a few simple potential cases such as the Coulomb, the
harmonic oscillator, the pseudoharmonic, isotonic potentials and others
[1-4]. The most interesting and best known system inside this small family
is the harmonic oscillator whose energy spectrum consists of an infinite set
of equidistant energy levels. Many other oscillators, as for example
harmonic oscillators perturbed by a term containing a fourth or a sixth
power in the coordinate, have been extensively studied. Nevertheless, it is
known the existence of another solvable one-dimensional model which shares
several interesting properties with the harmonic oscillator, the so-called
isotonic oscillator whose spectrum coincides with that of the harmonic
oscillator [5]. This potential model is considered as one of the most used
models for the study of the dynamics of nonlinear systems. In particular,
the energy spectrum of the singular potential is considerd as an isomorphous
to the harmonic oscillator spectrum. Very recently, Fellows and Smith [6]
have used the ideas of the factorization and supersymmetric (SUSY) approach
[7,8] to study the singular superpotentials and they proved that most of
these oscillatory potentials are other partner potentials of the harmonic
oscillators and derived an infinite set of exactly soluble potentials. A
discussion of the supersymmetric connection between harmonic and isotonic
oscillators can be traced in Ref. [9].

The isotonic potential takes the form,%
\begin{equation}
U_{Isot}(x)=U_{0}(x)+U_{g}(x)=\frac{1}{2}M\omega ^{2}x^{2}+\frac{1}{2}\frac{g%
}{x^{2}},\text{ }x\neq 0,\text{ }g\geq 0,
\end{equation}%
where $\omega $ is the angular frequency of oscillator, $M\omega ^{2}=K$ in
classical mechanics and $g=m(m+1).$ The common feature of this potential is
that it consists of a harmonic term plus an additional rational function
(centripetal barrier) which falls off at infinity like a constant $g$ times $%
1/x^{2}$ with one regular singularity at $x=0$ along the whole domain $%
-\infty <x<\infty .$\tablenotemark[1]%
\tablenotetext[1]{Centrifugal barrier
does not make physical sense in one-dimension. It is often used for such
singular terms in a potential.} It is clear from (1) that $U_{g}(x)$
exhibits strong singularity when $x=0,$ so that the wave functions must
vanish at such a point. It is worthy to note that the Hilbert space
associated to the models with $U_{g}(x)$ is narrowed as compared to the
Hilbert space of the hamiltonian with potential $U_{0}(x).$ In addition, the
domain of the hamiltonian associated to harmonic oscillator extends along
the whole real axis $\left( -\infty ,\infty \right) ,$ however, for
potential (1) reduces itself to the half-line $(x\geq 0$ or $x\leq 0).$ The
harmonic oscillator $U_{0}(x)$ and the isotonic oscillator $U_{Isot}(x)$ are
plotted in Figure 1 for the sake of comparison. The aim is to show that both
curves coincide for wide range of $x>0$ except for values of $x$ in the
neighborhood of the vertical asymptotic line $x=0$ where the isotonic
oscillator goes to infinity. They have also identical spectrum, however,
shifted by two units. We consider this potential in the interval $(0,\infty
) $ as in the case of (1). Authors of Refs. [7-9] have solved the Schr\"{o}%
dinger equation for the potential (1) with $\omega =M=1$ and $g=2$ $(m=1)$
using SUSY approach [7,8] on the assumption that the superpotential $W(x)$
obtained from the wave function of the harmonic oscillator $\phi _{1}(x)$
can generate twice the potential (1) (see e.g., [6])$.$ The singular term in
the above potential is often called the centripetal barrier potential.
However, the centripetal barrier potential here makes no physical sense in
one dimension since the term $m(m+1)/x^{2}$ singularities are often related
to the radial equation for the three-dimensional harmonic oscillator.

In this article we set up to present a study of the exact analytic
nonrelativistic bound state energy spectrum and the correspoding wave
functions in terms of the associated Laguerre polynomials $L_{n}^{\alpha
}(z) $ (or the Kummer confluent hypergeometric function, $M(a,b,z)=%
\begin{array}{c}
_{1}F_{1}%
\end{array}%
\left( a;b;z\right) $) by applying the Nikiforov-Uvarov (NU) method [10].
Overmore, we extend our study to investigate this potential model in the
context of the spin and pseudospin symmetric Dirac equation [11-17]. In the
presence of the spin symmetry $S\sim V=U_{Isot}(x)$ and pseudospin symmetry $%
S\sim -V=U_{Isot}(x)$, we investigate the exact $s$-wave bound state energy
eigenvalues and corresponding upper and lower spinor wave functions in a
systematic form [12-15]. We also show that the spin (pseudospin) symmetric
Dirac solutions can be reduced to the $S=V=U_{Isot}(x)$ ($S=-V=U_{Isot}(x))$
in the presence of exact spin symmetry $\Delta =0$ (pseudospin symmetry $%
\Sigma =0)$ limitation [16,17]. Overmore, the solution of the Dirac equation
can be easily reduced to it's nonrelativistic limit if one applies an
appropriate map of parameters.

The rest of the paper is organized as follows. In section 2, we apply the NU
method to solving the Schr\"{o}dinger and Dirac equations with an exactly
solvable isotonic oscillator to obtain the eigenvalues and eigenfunctions in
a systematical way. We also compare the non-relativistic solution with the
existing one obtained by applying the SUSY approach. In section 3, we make
our summary and conclusions.

\section{Bound State Solutions}

\subsection{Schr\"{o}dinger case}

We start with the one-dimensional single-particle Schr\"{o}dinger equation
[18,19]: 
\begin{equation}
H_{Isot}\psi _{n}(\pm x)=\left[ -\frac{\hbar ^{2}}{2M}\frac{d^{2}}{d(\pm
x)^{2}}+U_{Isot}(\pm x)\right] \psi _{n}(\pm x)=E_{n}\psi _{n}(\pm x),
\end{equation}%
The potential in Eq. (1) is invariant with respect to the inversion, i.e., $%
U_{Isot}(-x)\rightarrow U_{Isot}(x)$ as well as the hamiltonian $H_{Isot}.$
Therefore, the Schr\"{o}dinger equation should have even and odd solutions.
The domain of the harmonic oscillator hamiltonian $H$ extends along the
whole real axis $-\infty <x<\infty ,$ however, the partner $H_{Isot}$
exhibits a strong singularity at the origin. The space breaks up into two
disjoint regions $(x\geq 0$ or $x\leq 0)$ without communication between them
since the wave functions vanish at the regular singularity $x=0$ $($i.e., $%
\psi _{n}(0)=0)$ and at the irregular singularities $\pm \infty $ $($i.e., $%
\psi _{n}(\pm \infty )=0).$ In this respect, we should restrict the
hamiltonian to the interval $(0,\infty ),$ this is exactly the same
situation that occurs in isotonic potential. In addition, if we set $%
x\rightarrow ix,$ then (2) becomes%
\begin{equation*}
\left[ -\frac{\hbar ^{2}}{2M}\frac{d^{2}}{dx^{2}}+U_{Isot}(x)\right] \psi
_{n}(ix)=-E_{n}\psi _{n}(ix),
\end{equation*}%
which is the original equation with the irrelevant change in the eigenvalues 
$E_{n}\rightarrow -E_{n}.$ If we set $x\rightarrow ix$ in the isotonic wave
function, we get perfectly good $\psi _{n}(ix)$ which can generate the
superpotential $W(x)$ and other hamiltonians [6].

To solve Eq. (2) by NU method, we perform a straightforward algebra to
reduce it into the following simple form: 
\begin{equation}
\psi _{n}^{\prime \prime }(x)+\left[ \varepsilon _{n}-\beta ^{2}x^{2}-\frac{%
\alpha }{x^{2}}\right] \psi _{n}(x)=0,
\end{equation}%
with%
\begin{equation}
\varepsilon _{n}=\frac{2ME_{n}}{\hbar ^{2}},\text{ }\beta =\frac{M\omega }{%
\hbar },\text{ }\alpha =\frac{Mg}{\hbar ^{2}}\geq 0.
\end{equation}%
Let us restrict ourselves to the positive half-line $(x\geq 0)$ and in terms
of new variable $s=x^{2}$ $(0\leq s<\infty ),$ we obtain 
\begin{equation}
\psi _{n}^{\prime \prime }(s)+\frac{1}{\left( 2s\right) }\psi _{n}^{\prime
}(s)+\frac{1}{(2s)^{2}}\left[ -\beta ^{2}s^{2}+\varepsilon _{n}s-\alpha %
\right] \psi _{n}(s)=0,\text{ }\psi _{n}(0)=0,
\end{equation}%
where we have used $\psi _{n}(x)=\psi _{n}(s).$ Now, if we compare the above
equation with the following generalized hypergeometric-type equation with a
parametrization of real variables $s=s(x)$:%
\begin{equation}
\psi _{n}^{\prime \prime }(s)+\frac{\widetilde{\tau }(s)}{\sigma (s)}\psi
_{n}^{\prime }(s)+\frac{\widetilde{\sigma }(s)}{\sigma ^{2}(s)}\psi
_{n}(s)=0,
\end{equation}%
where%
\begin{equation}
\psi _{n}(s)=\Omega (s)y_{n}(s),
\end{equation}%
and where $\sigma (s)$ and $\widetilde{\sigma }(s)$ are two polynomials, at
most of second-degree, and $\widetilde{\tau }(s)$ is at most of first-degree
polynomial, then it follows that$:$ 
\begin{equation}
\widetilde{\tau }(s)=1,\text{ }\sigma (s)=2s,\text{ }\widetilde{\sigma }%
(s)=-\beta ^{2}s^{2}+\varepsilon _{n}s-\alpha .
\end{equation}%
To apply the NU method [10,20], we calculate the function $\pi (s)$ defined
by 
\begin{equation*}
\pi (s)=\frac{\sigma ^{\prime }(s)-\widetilde{\tau }(s)}{2}\pm \sqrt{\left[ 
\frac{\sigma ^{\prime }(s)-\widetilde{\tau }(s)}{2}\right] ^{2}-\widetilde{%
\sigma }(s)+k\sigma (s)}
\end{equation*}%
\begin{equation}
=\frac{1}{2}\left( 1\pm \sqrt{4\beta ^{2}s^{2}+4\left( 2k-\text{ }%
\varepsilon _{n}\right) s+4\alpha +1}\right) ,
\end{equation}%
and also seek for a physical value of $k$ that makes the discriminant of the
expression under square root, in the last equation, to become zero $($i.e., $%
2k=\varepsilon _{n}\pm \beta \sqrt{1+4\alpha },$ $\alpha \geq -1/4)$. Hence,
there is no bound solution in the region $\alpha <-1/4.$ The model becomes
unphysical in the region $(-\infty ,-1/4),$ since the spectrum is not
bounded from below. Upon the substitution of the value of $k$ into the above
equation, we obtain the following suitable solutions:%
\begin{equation}
\pi (s)=\frac{1}{2}\left( 1+\sqrt{1+4\alpha }\right) -\beta s,
\end{equation}%
and%
\begin{equation}
k=\frac{1}{2}\left( \varepsilon _{n}-\beta \sqrt{1+4\alpha }\right) .
\end{equation}%
With regard to Eqs. (8) and (10), we can calculate the function $\tau (s)=%
\widetilde{\tau }(s)+2\pi (s),$ taking into consideration the bound state
condition which has to be established when $\tau ^{\prime }(z)<0,$ as%
\begin{equation}
\tau (s)=2+\sqrt{1+4\alpha }-2\beta s\text{ \ and \ }\tau ^{\prime
}(s)=-2\beta <0.
\end{equation}%
According to the method, in order to find the energy equation from which one
calculates the energy eigenvalues, we need to find the values of the
parameters: $\lambdabar =k+\pi ^{\prime }(s)$ and $\lambdabar =\lambdabar
_{n}=-n\tau ^{\prime }(s)-\frac{1}{2}n\left( n-1\right) \sigma ^{\prime
\prime }(s),\ n=0,1,2,\cdots ,$ as%
\begin{equation}
\lambdabar =\frac{1}{2}\left( \varepsilon _{n}-\beta \sqrt{1+4\alpha }%
\right) -\beta \text{ and }\lambdabar _{n}=2n\beta ,\text{ }n=0,1,2,\cdots .
\end{equation}%
Using the relation $\lambdabar =\lambdabar _{n}$ and the definitions of
parameters in Eq. (4), one finds that the energy eigenvalues of the isotonic
oscillator are%
\begin{equation}
E_{n,g}=\hbar \omega \left( 2n+1+\frac{1}{2}\sqrt{1+\frac{4Mg}{\hbar ^{2}}}%
\right) ,\text{ }n=0,1,2,\cdots ,
\end{equation}%
which is identical to the results of Ref. [21] (see p. 3). Thus, we have
exactly solved the isotonic Hamiltonian (2). \ In the limit that $%
g\rightarrow 0,$ the relation (14) reduces to $E_{n}=\hbar \omega \left( 2n+%
\frac{3}{2}\right) ,$ $n=0,1,2,\cdots $ which is identical to the $s$-wave
solution of the three-dimensional Schr\"{o}dinger equation with harmonic
oscillator potential (cf. Eq. (34) when $l=0$).

Overmore, using the conventions of Ref. [9] (cf. Eqs. (41) and (43)
therein), we may take $g=2$ $($i.e., $m=1)$ for easy of notation, the
spectrum of Eq. (1) (in $\hbar =M=\omega =1$ units) reads as%
\begin{equation}
E_{n}=2n+\frac{5}{2},\text{ }n=0,1,2,\cdots .
\end{equation}%
It is noticed that energy spectrum in the previous equation is half the
spectrum of Eq. (41) in Ref. [9] (see Eq. (43) in [9]). The odd solutions
under the inversion $x\rightarrow -x$ (the negative half-line, $x\leq 0$)
has same energy spectrum as the even ones given in Eq. (14) due to the
invariance of isotonic potential under this inversion. Hence, the energy
relation in Eq. (14) holds for the whole domain $-\infty <x<\infty .$

Thus, all the other eigenenergies are given by 
\begin{equation}
E_{2n}=E_{0}+2n\omega ,\text{ }n=0,1,2,\cdots ,
\end{equation}%
and the energy spectrum is equidistant since%
\begin{equation}
E_{2n+2}=E_{2n}+2\omega .
\end{equation}%
Nevertheless, the height $\Delta E=2\omega $ of the energy steps is twice
that of the simple harmonic oscillator $U_{0}.$ In fact, it seems as if half
of of the states (those with an odd number of nodes) have disappeared.

Let us now turn to the calculations of the normalized wave function. The
first part of the wave function in Eq. (7) is found through the relation
[10,20]: 
\begin{equation}
\Omega (s)=\exp \left( \dint \frac{\pi (s)}{\sigma (s)}ds\right) =s^{\frac{1%
}{4}+\frac{1}{2}\xi }\exp \left( -\frac{1}{2}\beta s\right) ,
\end{equation}%
with%
\begin{equation}
\xi =\frac{1}{2}\sqrt{1+4\alpha }\geq \frac{1}{2},
\end{equation}%
and the calculation of the weight function is found through the relation,%
\begin{equation}
\rho (s)=\frac{1}{\sigma (s)}\exp \left( \dint \frac{\tau (s)}{\sigma (s)}%
ds\right) =s^{\xi }\exp \left( -\beta s\right) ,
\end{equation}%
leading to the calculation of the other part$\ $of the wave function;
namely,\ $y_{n}(s)$ which is a hypergeometric type function whose polynomial
solutions are given by the Rodrigues relation:%
\begin{equation}
y_{n}(s)=A_{n}\rho ^{-1}(s)\frac{d^{n}}{ds^{n}}\left[ \sigma ^{n}(s)\rho (s)%
\right] =s^{-\xi }\exp \left( \beta s\right) \frac{d^{n}}{ds^{n}}\left(
s^{n+\xi }\exp \left( -\beta s\right) \right) =L_{n}^{(\xi )}(\beta s),
\end{equation}%
:where $L_{n}^{\alpha }(y)$ is the associated Laguerre polynomials.
Therefore, the even solution of the wave function satisfying Eq. (7) is
[22,23] 
\begin{equation}
\psi _{n}(x)=\sqrt{\frac{2\beta ^{1+\xi }n!}{\Gamma \left( n+\xi +1\right) }}%
x^{\frac{1}{2}+\xi }\exp \left( -\frac{1}{2}\beta x^{2}\right) L_{n}^{(\xi
)}(\beta x^{2}),\text{ }\func{Re}\xi >0.
\end{equation}%
It should be noticed that the change of $x\rightarrow ix$ in the eigenvalue
equation (2) for the isotonic oscillator changes eigenvalues $%
E_{n}\rightarrow -E_{n}.$ Then, if $\psi _{n}(x)$ is the eigenfunction
corresponding to the eigenvalue $E_{n},$ then the eigenfunction $\psi
_{n}(ix)$ will be normalizable only if $-E_{n}$ is in the point spectrum of
this Hamiltonian. However, the $\psi _{n}(ix)$ would not be good wave
functions if their spectrum $-E_{n}$ lies not within the Hamiltonian range,
then $\psi _{n}(ix)$ is not normalizable eigenfunction$.$ At first, it
behaves like $\exp \left( \frac{1}{2}\beta x^{2}\right) $ at large $x$ but
this is not relevant here as one writes the explicit form of $L_{n}^{(\xi
)}(\beta x^{2})$ for even $n$. On the other hand, the odd solutions for
which the part of wave function corresponding to $-x$ have opposite signs
and exist as%
\begin{equation}
\psi _{n}(-x)=\mathcal{N}_{n}\left( -x\right) ^{\frac{1}{2}-\xi }\exp \left(
-\frac{1}{2}\beta x^{2}\right) L_{n}^{(\xi )}(\beta x^{2}).
\end{equation}%
The two linearly independent solutions (wave functions) given by Eqs. (22)
and (23) for the even and odd solutions, respectively, need to be
normalizable in the range $(0,\infty ).$ However$,$ the odd solution is not
normalizable in the region $(0,\infty )$ as one can see in Eq. (23)$.$
Indeed, the operator in (3) is not essentially self-adjoint for $-1/4\leq
\alpha <3/4$ and its most general square-integrable solution behaves near
the singularity as a linear combination of $x^{1/2+\xi }$ \ and $x^{1/2-\xi
} $ (See Eqs. (22) and (23)).\tablenotemark[1]%
\tablenotetext[1]{We would like to
thank one of the referees for drawing our attention to this point.} In this
respect, we clarify this point by analyzing the behaviour of the isotonic
potential (1) in terms of the parameter $\alpha $ [24,25], three different
regions appear, namely,

\begin{itemize}
\item In the range $\alpha \in (-\infty ,-1/4)$ the model becomes unphysical
since the spectrum is not bounded from below (see Figure 1) [24].

\item When $\alpha $ $\in (-1/4,3/4),$ the singularity is not strong enough
to make the wave functions (22) and (23) vanish at $x=0.$ Indeed in this
region both linearly independent solutions (wave functions) are normalizable
since $\xi =0$ and $1$. This is the reason why it is necessary to select,
from the continuous family of self-adjoint extensions by the differential
operator, the self-adjoint extensions which correctly describes the physical
system under consideration [24]. The wave functions pass across the
singularity point $x=0$ and the model extends itself again along the entire
region, that is; $\left( -\infty ,\infty \right) .$

\item Physically, if we consider the range $\alpha $ $\in (3/4,\infty ),$
the singularity acts as an impentrable barrier, thus dividing the space into
two independent regions, that is; $x\leq 0$ and $x>0.$ The wave functions
must vanish at $x=0$ which provides an absolute lack of communication
between the two regions of space (i.e., the negative and positive
half-lines) and the wave functions (22) and (23) in this case become
normalizable [24,25].
\end{itemize}

Alternatively, notice the Laguerre polynomial can be expressed in terms of
the Kummer confluent hypergeometric functions as [23]%
\begin{equation}
L_{n}^{p}(z)=\frac{\left( p+n\right) !}{p!n!}%
\begin{array}{c}
_{1}F_{1}%
\end{array}%
\left( -n;p+1;z\right) ),
\end{equation}%
where%
\begin{equation}
\begin{array}{c}
_{1}F_{1}%
\end{array}%
\left( a;b;z\right) =1+\frac{a}{b}z+\frac{a(1+a)}{2b(1+b)}z^{2}+\frac{%
a(1+a)(2+a)}{6b(1+b)(2+b)}z^{3}+O[z]^{4}.
\end{equation}%
Using the notations of other authors ($\hbar =M=1$)[21] and putting $\beta
=\omega $ and $\xi =m+1/2,$ the even wave function solution in Eq. (22)
becomes [23] 
\begin{equation*}
\psi _{n}(x)=N_{n}x^{1+m}\exp \left( -\frac{1}{2}\omega x^{2}\right) 
\begin{array}{c}
_{1}F_{1}%
\end{array}%
\left( -n;m+\frac{3}{2};\omega x^{2}\right) ,\text{ }n=0,2,4,\cdots ,
\end{equation*}%
\begin{equation}
N_{n}=\frac{1}{\Gamma \left( m+\frac{3}{2}\right) }\sqrt{\frac{2\omega ^{m+%
\frac{3}{2}}\Gamma \left( n+m+\frac{3}{2}\right) }{n!}},
\end{equation}%
which is identical to Eq. (42) in Ref. [9] when we set $m=1$ so that the
isotonic potential given by Eq. (41) in [9] is twice the potential (1) in
the present work. On the other hand, the odd solution for which the part of
wave function corresponding to $-x$ can be expressed as%
\begin{equation}
\psi _{n}(-x)=(-1)^{1+m}\psi _{n}(x),\text{ }n=1,3,5,\cdots .
\end{equation}%
Hence, if we take $m=0,2,4,\cdots $ (even real integer), we find the wave
function $\psi _{n}(-x)$ being an odd (antisymmetric) function of $x$ [i.e., 
$\psi _{n}(-x)=-\psi _{n}(x)].$ Overmore, if we take $m=1,3,5,\cdots $ (odd
real integer), we find the wave function corresponding to negative values of 
$x$ is identical to the wave function corresponding to positive values,
i.e., $\psi _{n}(-x)$ being an even function (symmetric) with $\psi
_{n}(-x)=\psi _{n}(x).$ In case if $m$ is a real number but not integer
yielding a complex wave function in the negative half-line which is not
normalizable.

On the other hand, the energy levels of the one-dimensional Schr\"{o}dinger
equation for the harmonic oscillator $U_{0}(x)$ [1]:%
\begin{equation}
E_{n}=\left( n+\frac{1}{2}\right) \hbar \omega ,\text{ }n=0,1,2,\cdots ,
\end{equation}%
and the well-known wave functions [26]%
\begin{equation}
\text{ }\phi _{n}(x)=\left[ \frac{1}{2^{n}n!}\sqrt{\frac{\beta }{\pi }}%
\right] ^{1/2}H_{n}(\beta x^{2})\exp (-\frac{1}{2}\beta x^{2}),\text{ }\beta
=\frac{M\omega }{\hbar },
\end{equation}%
where $H_{n}(y)=(-1)^{n}\exp (y^{2})\frac{d^{n}}{dy^{n}}\exp (-y^{2})$
represent Hermite polynomials.

For further discussions on the isotonic potential, we present the energy
eigenvalues and the corresponding wave functions of the lowest three states, 
\begin{subequations}
\begin{equation}
E_{0}=\omega \left( \frac{3}{2}+m\right) ,
\end{equation}%
\begin{equation}
\psi _{0}(x)=N_{0}x^{1+m}\exp \left( -\frac{1}{2}\omega x^{2}\right) ,
\end{equation}%
\end{subequations}
\begin{subequations}
\begin{equation}
E_{1}=\left( \frac{5}{2}+m\right) \omega ,
\end{equation}%
\begin{equation}
\psi _{1}(x)=N_{1}x^{1+m}\exp \left( -\frac{1}{2}\omega x^{2}\right) \left(
1-\frac{2\omega }{\left( 2m+3\right) }x^{2}\right) ,
\end{equation}%
and 
\end{subequations}
\begin{subequations}
\begin{equation}
E_{2}=\left( \frac{7}{2}+m\right) \omega ,
\end{equation}%
\begin{equation}
\psi _{2}(x)=N_{2}x^{1+m}\exp \left( -\frac{1}{2}\omega x^{2}\right) \left(
1-\frac{4\omega }{\left( 2m+3\right) }x^{2}+\frac{4\omega ^{2}}{\left(
2m+3\right) \left( 2m+5\right) }x^{4}\right) ,
\end{equation}%
respectively and the normalization factors are calculated as 
\end{subequations}
\begin{equation}
N_{0}=\sqrt{\frac{2\omega ^{m+\frac{3}{2}}}{\Gamma \left( m+\frac{3}{2}%
\right) }},\text{ }N_{2}=(-1)^{1+m}\sqrt{\frac{\omega ^{m+\frac{3}{2}}\left(
2m+3\right) }{\Gamma \left( m+\frac{3}{2}\right) }},\text{ }N_{2}\sqrt{\frac{%
\omega ^{m+\frac{3}{2}}\left( 2m+5\right) \left( 2m+3\right) }{4\Gamma
\left( m+\frac{3}{2}\right) }}.
\end{equation}%
The isotonic ground state wave function, $\psi _{0}(x)$ in (30b) is compared
with the corresponding harmonic oscillator wave function, $\phi _{0}(x)$ in
Eq. (29) in Figure 2. Further, the isotonic first two excited wave
functions, $\psi _{1}(x)$ and $\psi _{2}(x)$ in Eqs. (31b) and (32b) are
compared with the corresponding harmonic oscillator wave functions, $\phi
_{1}(x)$\ and $\phi _{2}(x)$ in Eq. (29) in Figures 3 and 4, respectively.

On the other hand, the solution of the three-dimensional Schr\"{o}dinger
equation with any arbitrary quantum number $l$ \ (i.e., harmonic oscillator
combined with centrifugal barrier potential) provides us%
\begin{equation}
E_{n,l}=\hbar \omega \left( 2n+l+\frac{3}{2}\right) ,\text{ }n,l=0,1,2,\cdots
\end{equation}%
and the corresponding wave functions are given by%
\begin{equation}
\text{ }\psi _{n,l}(r,\theta ,\varphi )=\sqrt{\left( \frac{\beta }{\pi }%
\right) ^{1/2}\frac{2^{n+2l+3}n!\left( 2\beta \right) ^{l}}{\left(
2n+2l+1\right) !!}}r^{l}\exp (-\frac{1}{2}\beta r^{2})L_{n}^{\left(
l+1/2\right) }\left( \beta r^{2}\right) Y_{l,m}(\theta ,\varphi ),
\end{equation}%
where $L_{n}^{\left( l+1/2\right) }\left( \beta r^{2}\right) $ is the
associated Laguerre polynomial, and $Y_{l,m}(\theta ,\varphi )$ is the
angular part of the wave functions. The order $n$ of the polynomial is a
non-negative integer. Thus, the exact solution of the isotonic oscillator in
Eq. (14) in one-dimension is equivalent to the solution of the harmonic
oscillator $U_{0}(r)$ combined with the centrifugal barrier potential $%
l(l+1)/r^{2},$ $r\in (0,\infty ),$ in three-dimensions given in Eq. (34)
when we take $\hbar =M=1$ and $g=m(m+1),$ where $g$ is a real number. \ That
is, $E_{n,m}=\hbar \omega \left( n_{1}+\frac{3}{2}\right) $ is equivalent to 
$E_{n,l}=\hbar \omega \left( n_{2}+\frac{3}{2}\right) ,$ where we have
defined $n_{1}=2n+m$ and $n_{2}=2n+l$ ($m\leftrightarrow l$) for which the
solutions are defined for positive half-line [$x\in (0,\infty
)\leftrightarrow r\in (0,\infty )$]$.$

\subsection{Dirac Case}

We start by writting the two radial coupled Dirac equations for the upper
and lower (i.e., $F_{n,\kappa }(r)$ and $G_{n,\kappa }(r),$ respectively$)$
spinor components [26,27]: 
\begin{subequations}
\begin{equation}
\left( \frac{d}{dr}+\frac{\kappa }{r}\right) F_{n,\kappa }(r)=\left(
Mc^{2}+E_{n\kappa }-\Delta \right) G_{n,\kappa }(r),
\end{equation}%
\begin{equation}
\left( \frac{d}{dr}-\frac{\kappa }{r}\right) G_{n,\kappa }(r)=\left(
Mc^{2}-E_{n\kappa }+\Sigma \right) F_{n,\kappa }(r),
\end{equation}%
where $\Delta =V-S$ and $\Sigma =V+S$ are the difference and sum potentials,
respectively and are expressed in terms of vector ($V)$ and scalar ($S)$
potentials. In addition, $c\approx 137$ is the velocity of light.

In the presence of spin symmetry ( i.e., $\Delta =C_{s}$), one gets a
second-order differential equation satisfying the upper-spinor component
[17,28-31] 
\end{subequations}
\begin{equation}
F_{n\kappa }^{\prime \prime }(r)-\left( \frac{\kappa \left( \kappa +1\right) 
}{r^{2}}+A_{s}^{2}+\gamma \Sigma \right) F_{n\kappa }(r)=0,
\end{equation}%
where 
\begin{equation}
A_{s}^{2}=\gamma \left( Mc^{2}-E_{n\kappa }\right) ,\text{ }\gamma =\frac{1}{%
\hbar ^{2}c^{2}}\left( Mc^{2}+E_{n\kappa }-C_{s}\right) >0,
\end{equation}%
and $\kappa \left( \kappa +1\right) =l\left( l+1\right) ,$ $\kappa =l$ for $%
\kappa <0$ and $\kappa =-\left( l+1\right) $ for $\kappa >0.$ The spin
symmetry energy eigenvalues depend on $n$ and $\kappa ,$ \textit{i.e.}, $%
E_{n\kappa }=E(n,\kappa \left( \kappa +1\right) ).$ For $l\neq 0,$ the
states with $j=l\pm 1/2$ are degenerate. Further, the lower-spinor component
can be obtained from Eq. (36a) as 
\begin{equation}
G_{n\kappa }(r)=\frac{1}{Mc^{2}+E_{n\kappa }-C_{s}}\left( \frac{d}{dr}+\frac{%
\kappa }{r}\right) F_{n\kappa }(r),
\end{equation}%
where $E_{n\kappa }\neq -Mc^{2},$ i.e., only real positive energy states
exist when $C_{s}=0$ (exact spin symmetric case).

On the other hand, under the pseudospin symmetry ( i.e., $\Sigma =C_{ps}$),
one obtains a second-order differential equation satisfying the lower-spinor
component,%
\begin{equation}
G_{n\kappa }^{\prime \prime }(r)-\left( \frac{\kappa \left( \kappa -1\right) 
}{r^{2}}+A_{ps}^{2}-\widetilde{\gamma }\Delta \right) G_{n\kappa }(r)=0,
\end{equation}%
where 
\begin{equation}
A_{ps}^{2}=\widetilde{\gamma }\left( Mc^{2}+E_{n\kappa }\right) ,\text{ }%
\widetilde{\gamma }=\frac{1}{\hbar ^{2}c^{2}}\left( Mc^{2}-E_{n\kappa
}+C_{ps}\right) ,
\end{equation}%
and the upper-spinor component $F_{n\kappa }(r)$ is obtained from Eq. (36b)
as 
\begin{equation}
F_{n\kappa }(r)=\frac{1}{Mc^{2}-E_{n\kappa }+C_{ps}}\left( \frac{d}{dr}-%
\frac{\kappa }{r}\right) G_{n\kappa }(r),
\end{equation}%
where $E_{n\kappa }\neq Mc^{2},$ i.e., only real negative energy states
exist when $C_{ps}=0$ (exact pseudospin symmetric case). From the above
equations, the energy eigenvalues depend on the quantum numbers $n$ and $%
\kappa $, and also the pseudo-orbital angular quantum number $\widetilde{l}$
according to $\kappa (\kappa -1)=\widetilde{l}(\widetilde{l}+1),$ which
implies that $j=\widetilde{l}\pm 1/2$ are degenerate for $\widetilde{l}\neq
0.$ The quantum condition for bound states demands the finiteness of the
solution at infinity and at the origin points, i.e., $F_{n\kappa
}(0)=G_{n\kappa }(0)=0$ and $F_{n\kappa }(\infty )=G_{n\kappa }(\infty )=0.$

Let us now study the isotonic potential (1) in the context of spin and
pseudospin symmetric Dirac equations. It is well-known that Eqs. (37) and
(40) can be solved exactly for any $\kappa $ with the spin-orbit
(pseudospin-orbit) centrifugal (pseudo centrifugal) potential term. However,
we shall study these equations for the $s$-wave case ($\kappa =\pm 1$) for
the sake of comparison with the nonrelativistic case since $m(m+1)/x^{2}$ in
the isotonic potential has the same behaviour as $\kappa (\kappa \pm
1)/r^{2} $ in Eqs. (37) and (40)$.$

\subsubsection{Spin symmetry limit}

This symmetry arises from the near equality in magnitude of an attractive
scalar, $S,$ and repulsive vector, $V,$ relativistic mean field, $S\sim V$
in which the nucleon move [12]. Therefore, we simply take the sum potential
equal to the isotonic potential model, i.e.,%
\begin{equation}
\Sigma =U_{Isot}(x)=\frac{1}{2}M\omega ^{2}x^{2}+\frac{1}{2}\frac{g}{x^{2}}.
\end{equation}%
In the last equation, the choice of $\Sigma =2V\rightarrow U_{Isot}(x)$ as
stated in Ref. [26] allows one to reduce it into its non-relativistic limit
under appropriate choice of parameter transformations. Further, we take $%
\kappa =-1$ ($l=0$) and in terms of new variable $s=x^{2}$ $($positive
half-plane $x\geq 0),$ Eq. (37) becomes%
\begin{equation}
F_{n,-1}^{\prime \prime }(s)+\frac{1}{(2s)}F_{n,-1}^{\prime }(s)+\frac{1}{%
(2s)^{2}}\left[ -\nu ^{2}s^{2}-A_{s}^{2}s-\beta \right] F_{n,-1}(s)=0,
\end{equation}%
where 
\begin{equation}
\beta =\frac{1}{2}g\gamma \text{ \ and \ }\nu =\sqrt{\frac{1}{2}M\omega
^{2}\gamma }.
\end{equation}%
The quantum condition is obtained from the finiteness of the solution at
infinity and at the origin point$.$ We apply the NU method following the
same steps of solution in previous section to obtain the expressions: 
\begin{equation}
\widetilde{\tau }(s)=1,\text{\ }\sigma (s)=2s,\text{\ }\widetilde{\sigma }%
(s)=-\nu ^{2}s^{2}-A_{s}^{2}s-\beta .
\end{equation}%
It follows that the functions required by the method for $\pi (s),$ $k$ and $%
\tau (s)$ take the suitable forms: 
\begin{equation}
\pi (s)=-\nu s+\frac{1}{2}\left( 1+\sqrt{1+4\beta }\right) ,
\end{equation}%
\begin{equation}
k=-\frac{1}{2}\left( A_{s}^{2}+\nu \sqrt{1+4\beta }\right) ,
\end{equation}%
and%
\begin{equation}
\tau (s)=2+\sqrt{1+4\beta }-2\nu s\text{ \ and \ }\tau ^{\prime }(s)=-2\nu
<0,
\end{equation}%
respectively, with prime denotes the derivative with respect to $s.$ Also,
the parameters $\lambdabar $ and $\lambdabar _{n}$ take the forms:%
\begin{equation}
\lambda =-\frac{1}{2}\left( A_{s}^{2}+\nu \sqrt{1+4\beta }\right) -\nu ,%
\text{ and }\lambda _{n}=2n\nu .
\end{equation}%
Using the condition $\lambdabar =\lambdabar _{n}$ followed by simple algebra$%
,$ we obtain the following transcendental energy equation,%
\begin{equation}
\left( E_{n,-1}-Mc^{2}\right) \sqrt{Mc^{2}+E_{n,-1}-C_{s}}=\hbar c\omega 
\sqrt{2M}\left( 2n+1+\frac{1}{2}\sqrt{\frac{2g}{\hbar ^{2}c^{2}}\left(
Mc^{2}+E_{n,-1}-C_{s}\right) +1}\right) ,
\end{equation}%
where $n=0,1,2,3,\cdots $ and $E_{n,-1}\geq C_{s}-Mc^{2}.$ One can compute
the energy eigenvalues by choosing suitable parameters in the symmetric
potential. Hence, Eq. (51) shows the energy eigenvalues $E_{n}$ dependence
on $n$ and $C_{s}$ as well as on the parameters $\omega $ and $M.$

Therefore, using Eq. (51), we compute some energy levels for several values
of $n$ (in units $\hbar =c=1$)$.$ In the presence spin symmetric limit,
Table 1 gives some numerical results by taking the following parameters
values: $M=$ $\omega =1.0$ $fm^{-1},$ $C_{s}=0$ $fm^{-1}$ (exact symmetric
case) and $C_{s}=2.0$ $fm^{-1}$ (non exact symmetric case). Moreover, the
strength of the centripetal barrier term is set up to some arbitrarily
chosen values: $g=0.5,$ 2 and $6$ corresponding to $m=0.3660254$, $1$ and $%
2, $ respectively. For the values $g=$ 2 and $6,$ the singularity acts as
impenetrable barrier, thus deviding the space into two independent regions,
the negative half-line and the positive half-line.

Dirac equation which in the limit of a non-relativistic and spinless
particle transforms into Schr\"{o}dinger equation for the isotonic potential
(1) is constructed as follows. In the exact spin symmetry, we set $C_{s}=0$
and apply appropriate transformations given by $\left(
Mc^{2}+E_{n,-1}\right) /\hbar ^{2}c^{2}\simeq 2M/\hbar ^{2}$ and $%
E_{n,-1}-Mc^{2}\simeq E_{n},$ we finally obtain the Schr\"{o}dinger solution
in (14).

Let us now turn to the calculations of the corresponding wave functions for
this system. We obtain the first part $\phi (s)$ of the wave function (7)
and the weight function $\rho (s)$ as%
\begin{equation}
\Omega (s)=s^{\frac{1}{4}1+\frac{1}{2}\zeta }\exp \left( -\frac{1}{2}\nu
s\right) ,
\end{equation}%
where%
\begin{equation}
\zeta =\frac{1}{2}\sqrt{1+2g\gamma },\text{ }g=m(m+1),
\end{equation}%
and 
\begin{equation}
\rho (s)=s^{\zeta }\exp \left( -\nu s\right) .
\end{equation}%
Hence, the second part $y_{n}(s)$ of the wave function (7) can be obtained
from the weight function as%
\begin{equation}
y_{n}(s)\sim L_{n}^{(\zeta )}\left( \nu s\right) .
\end{equation}%
Finally, we find the normalized wave function satisfying Eq. (37) as%
\begin{equation}
F_{n,-1}(x)=\sqrt{\frac{2\nu ^{1+\zeta }n!}{\Gamma \left( n+\zeta +1\right) }%
}x^{\frac{1}{2}+\zeta }\exp \left( -\frac{1}{2}\nu x^{2}\right)
L_{n}^{(\zeta )}\left( \nu x^{2}\right) .
\end{equation}%
In addition, the corresponding lower-spinor component wave function $%
G_{n,-1}(x)$ is found from the solution of Eq. (39) as%
\begin{equation*}
G_{n,-1}(x)=\frac{1}{\left( Mc^{2}+E_{n,-1}-C_{s}\right) }\sqrt{\frac{2\nu
^{1+\zeta }n!}{\Gamma \left( n+\zeta +1\right) }}x^{\frac{1}{2}+\zeta }\exp
\left( -\frac{1}{2}\nu x^{2}\right)
\end{equation*}%
\begin{equation}
\times \left[ \left( \frac{-1+2\zeta }{2x}-\nu x\right) L_{n}^{(\zeta
)}\left( \nu x^{2}\right) +\frac{dL_{n}^{(\zeta )}\left( \nu x^{2}\right) }{%
dx}\right] .
\end{equation}%
Let us remark that the obtained results of the $s$-wave ($\kappa =-1$) of
the spin-symmetric Dirac equation with isotonic oscillator agree with the
results of the three dimensional Dirac equation with harmonic oscillator
potential combined with the centrifugal barrier term $\kappa \left( \kappa
+1\right) /r^{2}$ if we make the parameter change $m\leftrightarrow \kappa $
since the positive half-line $x\in \left( 0,\infty \right) $ in the first is
also equivalent to $r\in \left( 0,\infty \right) $ in the second. This is
apparent because the centripetal barrier potential $m(m+1)/x^{2}$ in the
isotonic oscillator is equivalent to the centrifugal term $\kappa \left(
\kappa +1\right) /r^{2}$ in Eq. (37).

On the other hand, the Klein-Gordon solution for the isotonic potential (in
relativistic $\hbar =c=1$ units) can be obtained from the exact
spin-symmetric case, $V=S,$ $C_{s}=0.$ Hence, the energy equation can be
obtained from Eq. (51) as%
\begin{equation}
\left( E_{n,-1}^{2}-M^{2}\right) \left( E_{n,-1}-M\right) =2M\omega
^{2}\left( 2n+1+\frac{1}{2}\sqrt{1+2g\left( Mc^{2}+E_{n,-1}\right) }\right)
^{2},
\end{equation}%
and the wave function from Eq. (56) as%
\begin{equation*}
F_{n,-1}(x)=\sqrt{\frac{2n!\left( \epsilon \right) ^{\frac{1}{2}\left(
1+\lambda _{0}\right) }}{\Gamma \left( n+\lambda _{0}+1\right) }}x^{\frac{1}{%
2}+\lambda _{0}}\exp \left( -\frac{1}{2}\epsilon x^{2}\right)
L_{n}^{(\lambda _{0})}\left( \epsilon x^{2}\right) ,
\end{equation*}%
\begin{equation}
\epsilon =\sqrt{\frac{1}{2}M\omega ^{2}\left( M+E_{n,-1}\right) },\text{ }%
\lambda _{0}=\frac{1}{2}\sqrt{1+2g\left( M+E_{n,-1}\right) }.
\end{equation}

\subsubsection{Pseudospin symmetry limit}

The exact pseudospin symmetry occurs when $S\sim -V$ or $\Sigma =C_{ps}=$
constant [12,16] and the quality of the pseudospin approximation in real
nuclei is connected with the competition between the pseudo-centrifugal
barrier and the pseudospin-orbital potential [32]. Therefore, we take the
difference potential in Eq. (40) as the isotonic potential model, i.e.,%
\begin{equation}
\Delta =U_{Isot}(x)=\frac{1}{2}M\omega ^{2}x^{2}+\frac{1}{2}\frac{g}{x^{2}},
\end{equation}%
In the pseudospin symmetry, the eigenstates with with $\widetilde{j}=%
\widetilde{l}\pm \frac{1}{2}$ are degenerate for $\widetilde{l}\neq 0.$ For
the $s$-wave case ( $\kappa =1)$ and in terms of the variable $s=x^{2},$ Eq.
(40) reduces to a simple form%
\begin{equation}
G_{n,1}^{\prime \prime }(s)+\frac{1}{(2s)}G_{n,1}^{\prime }(s)+\frac{1}{%
(2s)^{2}}\left[ \widetilde{\nu }^{2}s^{2}-A_{ps}^{2}s+\widetilde{\beta }%
\right] G_{n,1}(s)=0,
\end{equation}%
where 
\begin{equation}
\widetilde{\beta }=\frac{1}{2}g\widetilde{\gamma }\text{ and \ }\widetilde{%
\nu }=\sqrt{\frac{1}{2}M\omega ^{2}\widetilde{\gamma }}.\text{ }
\end{equation}%
To avoid repetition in the solution of Eq. (61), a first inspection for the
relationship between the present set of parameters $(A_{ps}^{2},\widetilde{%
\beta },\widetilde{\nu })$ and the previous set $(A_{s}^{2},\beta ,\nu )$
provides that the energy solution for pseudospin symmetry can be similarly
found directly from those of the previous energy solutions for spin symmetry
using the following parameters map [33]: 
\begin{equation*}
F_{n,-1}(s)\leftrightarrow G_{n,1}(s),\text{ }E_{n,-1}\rightarrow -E_{n,1},%
\text{ }C_{s}\rightarrow -C_{ps},\text{ }A_{s}^{2}\rightarrow A_{ps}^{2}
\end{equation*}%
\begin{equation}
U(s)\rightarrow -U(s)\text{ }(\nu ^{2}\rightarrow -\widetilde{\nu }^{2},%
\text{ }\beta \rightarrow -\widetilde{\beta }\text{ or }\omega \rightarrow
j\omega ,\text{ }g\rightarrow -g),\text{ }j=\sqrt{-1},
\end{equation}%
from which trivial calculus gives us the transcendental energy equation:%
\begin{equation}
\left( E_{n,1}+Mc^{2}\right) \sqrt{E_{n,1}-Mc^{2}-C_{ps}}=\hbar c\omega 
\sqrt{2M}\left( 2n+1+\widetilde{\zeta }\right) ,
\end{equation}%
with%
\begin{equation}
\widetilde{\zeta }=\frac{1}{2}\sqrt{1+\frac{2g}{\hbar ^{2}c^{2}}\left(
E_{n,1}-Mc^{2}-C_{ps}\right) }.
\end{equation}%
where $n=0,1,2,3,\cdots $ and $E_{n,1}\geq Mc^{2}+C_{ps}$ is the main
condition for the real bound state solutions.

Now the normalized lower-spinor component wavefunctions of the isotonic
oscillator are given by%
\begin{equation}
F_{n,1}(x)=\sqrt{\frac{2\left( i\widetilde{\nu }\right) ^{1+\widetilde{\zeta 
}}n!}{\Gamma \left( n+\widetilde{\zeta }+1\right) }}x^{\frac{1}{2}+%
\widetilde{\zeta }}\exp \left( -\frac{1}{2}\widetilde{\nu }x^{2}\right)
L_{n}^{(\widetilde{\zeta })}\left( i\widetilde{\nu }x^{2}\right) .
\end{equation}%
Therefore, using Eq. (64), we compute some energy levels for several values
of $n.$ In the pseudospin symmetric limit, Table 2 gives some numerical
results by taking the following parameters values: $M=$ $\omega =1.0$ $%
fm^{-1},$ $C_{s}=0$ $fm^{-1}$ (exact symmetric case) and $C_{s}=-2.0$ $%
fm^{-1},$ $-13.0$ $fm^{-1}$ (non exact symmetric case). In addition, the
strength of the centripetal barrier term is set up to the following
arbitrarily chosen values: $g=0.5,$ 2$,$ and $6.$

\section{Conclusions and Outlook}

In this work, qualitative data were obtained on the modifications of
spectrum energy on a nonrelativistic and relativistic particle confined by
isotonic oscillator field of specific strength $g.$ The spin and pseudospin
symmetry in relativistic isotonic oscillator are investigated systemically
by solving the Dirac equation with scalar and vector radial potentials by
applying the NU method. In one-dimensional isotonic oscillator$,$ we have
obtained the exact solutions in closed form for the energy spectrum and the
wave functions, which are equivalent to solving the three-dimensional
harmonic oscillator problem. The isotonic oscillator is an isospectral to
harmonic oscillator. Also, the energy steps are twice that of the simple
harmonic oscillator. The resulting solutions of the wave functions are
written in terms of the associated Laguerre polynomials $L_{n}^{\alpha }(z)$
(confluent hypergeometric functions $M(a,b,z)=%
\begin{array}{c}
_{1}F_{1}%
\end{array}%
\left( a;b;z\right) $) and the wave function for states $n=0,1$ and $2$ are
found to have the same shape as the harmonic oscillator as shown in Figures
2, 3 and 4. The case where n is even appears to be the most interesting,
since all generated wave functions are normalizable. However, when $n$ is
odd, half of the generated wave functions must be removed as they are not
normalizable.

In the relativistic case, it is found that the solutions when $\Delta
=0\rightarrow S=V$\ (\textit{i.e}., exact symmetric case, $C_{s}=0$) or $%
\Sigma =0\rightarrow S=-V$\ (\textit{i.e}., exact pseudosymmetric case, $%
C_{ps}=0$) are identical to the Klein-Gordon solutions. Besides, they can be
readily reduced to the expected nonrelativistic limit when appropriate
mapping transformations of parameters are made. In the numerical work, the
relativistic energy spectrum for the spin and pseudospin symmetries are
given in Tables 1 and 2, respectively. It is noticed that the parameters $g,$
$M,$ $C_{s}$ and $C_{ps}$ should be adjusted to provide us real solutions
for the energy eigenvalues and eigenfunctions.

Finally, let us also mention that the isotonic oscillator possesses a
remarkable property. The change of $x\rightarrow ix$ in the wave equation
resulting in the change of eigenvalues $E_{n}\rightarrow -E_{n}.$ Then, if $%
\psi _{n}(x)$ is the eigenfunction corresponding to the eigenvalue $E_{n},$
then the eigenfunction $\psi _{n}(ix)$ will be normalizable only if $-E_{n}$
is in the point spectrum of this Hamiltonian. So the isotonic oscillator
wave function with the change $x\rightarrow ix$ would be a good wave
function as they are normalizable ($-E_{n}$ $\in \left\langle n\left\vert
H\right\vert n\right\rangle $) or would not be good if they are not
normalizable ($-E_{n}$ $\notin \left\langle n\left\vert H\right\vert
n\right\rangle $). The eigenfunction $\psi _{n}(ix)$ can be used to generate
new operators in the supersymmetric quantum mechanics [6,7]. This remains as
an open question that deserves to be studied.

\acknowledgments We acknowledge the kind referee(s) for the inavaluable
suggestions that helped us to improve this paper greatly.

\newpage

\ {\normalsize 
}

\bigskip

\baselineskip= 2\baselineskip
\bigskip \newpage

\bigskip

\bigskip {\normalsize 
}

\baselineskip= 2\baselineskip
\FRAME{ftbpFO}{0.0277in}{0.0277in}{0pt}{\Qct{Behaviour of the isotonic
oscillator potential (continuous line) and the harmonic oscillator potential
(dash line).}}{}{Figure 1}{}

\FRAME{ftbpFO}{0.0277in}{0.0277in}{0pt}{\Qct{Behaviour of the ground state
wave function $\protect\psi _{n=0,m=1}(x)$ of the isotonic oscillator
(continuous line) and the corresponding wave function $\protect\phi _{0}(x)$
of the harmonic oscillator (dash line).}}{}{Figure 2}{}\FRAME{ftbpFO}{%
0.0277in}{0.0277in}{0pt}{\Qct{Behaviour of the first excited wave function $%
\protect\psi _{n=1,m=1}(x)$ of the isotonic oscillator (continuous line) and
the corresponding wave function $\protect\phi _{1}(x)$ of the harmonic
oscillator (dash line).}}{}{Figure 3}{}\bigskip \FRAME{ftbpFO}{0.0277in}{%
0.0277in}{0pt}{\Qct{Behaviour of the second excited wave function $\protect%
\psi _{n=2,m=1}(x)$ of the isotonic oscillator (continuous line) and the
corresponding wave function $\protect\phi _{2}(x)$ of the harmonic
oscillator (dash line).}}{}{Figure 4}{}

\bigskip

\begin{table}[tbp]
\caption{The spin symmetric bound state energy eigenvalues (in $fm^{-1}),$
for several values of $n$ with parameter values $M=1.0$ $fm^{-1}$ and $%
\protect\omega =1.0$ $fm^{-1}.$ }%
\begin{tabular}{llllll}
\tableline\tableline & $C_{s}=0$ $fm^{-1}$ \tablenotetext[1]{Exact spin
symmetric limit.}\tablenotemark[1] &  &  & $C_{s}=2.0$ $fm^{-1}$ &  \\ 
$n/E_{n}$ & $g=0.5$ $\left( m\approx 0.366\right) $ & $g=2$ $(m=1)$ & $g=6$ $%
(m=2)$ & $g=2$ $(m=1)$ & $g=6$ $(m=2)$ \\ 
\tableline$0$ & $2.5509860$ & $3.1503636$ & $4.0959121$ & $3.3991120$ & $%
4.2634174$ \\ 
$1$ & $3.7292142$ & $4.2915849$ & $5.1735045$ & $4.6747397$ & $5.4772542$ \\ 
$2$ & $4.7223578$ & $5.2667833$ & $6.1147629$ & $5.7095838$ & $6.4867680$ \\ 
$3$ & $5.6093599$ & $6.1428129$ & $6.9690531$ & $6.6208542$ & $7.3835758$ \\ 
$4$ & $6.4244044$ & $6.9503157$ & $7.7611866$ & $7.4521361$ & $8.2052891$ \\ 
$5$ & $7.1861562$ & $7.7065008$ & $8.5058073$ & $8.2256717$ & $8.9719327$ \\ 
$6$ & $7.9061955$ & $8.4222280$ & $9.2124501$ & $8.9547327$ & $9.6957461$ \\ 
$7$ & $8.5923225$ & $9.1048960$ & $9.8877527$ & $9.6480343$ & $10.3848919$
\\ 
$8$ & $9.2501029$ & $9.7598277$ & $10.5365663$ & $10.3116853$ & $11.0451537$
\\ 
$9$ & $9.8836823$ & $10.3910117$ & $11.1625702$ & $10.9501754$ & $11.6808166$
\\ 
$10$ & $10.4962522$ & $11.0015335$ & $11.7686371$ & $11.5669263$ & $%
12.2951658$ \\ 
\tableline &  &  &  &  & 
\end{tabular}%
\end{table}
\begin{table}[tbp]
\caption{The pseudospin symmetric bound state energy eigenvalues (in $%
fm^{-1}),$ for several values of $n$ with parameter values $M=1.0$ $fm^{-1}$
and $\protect\omega =1.0$ $fm^{-1}.$ }%
\begin{tabular}{lllllllll}
\tableline\tableline & $C_{ps}=0$ \tablenotetext[1]{Exact pseudospin
symmetric limit.}\tablenotemark[1] &  &  & $C_{ps}=-2.0$ &  &  & $%
C_{ps}=-13.0$ &  \\ 
$n/E_{n}$ & $g=0.5$ & $g=2$ & $g=6$ & $g=0.5$ & $g=2$ & $g=6$ & $g=2$ & $g=6 
$ \\ 
\tableline$0$ & $1.7353829$ & $1.9975105$ & $2.6220370$ & $0.8996794$ & $%
1.3991120$ & $2.2634174$ & $0.8228652$ & $1.8370383$ \\ 
$1$ & $2.9274128$ & $3.2918405$ & $3.9528022$ & $2.1870188$ & $2.6747397$ & $%
3.4772541$ & $1.5785297$ & $2.5680523$ \\ 
$2$ & $3.9414440$ & $4.3370543$ & $5.0071893$ & $3.2260195$ & $3.7095838$ & $%
4.4867680$ & $2.2966386$ & $3.2659358$ \\ 
$3$ & $4.8433785$ & $5.2545579$ & $5.9290480$ & $4.1395244$ & $4.6208542$ & $%
5.3835758$ & $2.9834157$ & $3.9357442$ \\ 
$4$ & $5.6693464$ & $6.0900511$ & $6.7671403$ & $4.9722337$ & $5.4521361$ & $%
6.2052891$ & $3.6435022$ & $4.5813401$ \\ 
$5$ & $6.4394382$ & $6.8666546$ & $7.5454937$ & $5.7467734$ & $6.2256717$ & $%
6.9719327$ & $4.2804724$ & $5.2057558$ \\ 
$6$ & $7.1660777$ & $7.5980685$ & $8.2781774$ & $6.4765859$ & $6.9547326$ & $%
7.6957461$ & $4.8971501$ & $5.8114252$ \\ 
$7$ & $7.8575782$ & $8.2932428$ & $8.9743213$ & $7.1704749$ & $7.6480344$ & $%
8.3848919$ & $5.4958138$ & $6.4003383$ \\ 
$8$ & $8.5198335$ & $8.9584266$ & $9.6402732$ & $7.8345997$ & $8.3116853$ & $%
9.0451537$ & $6.0783346$ & $6.9741474$ \\ 
$9$ & $9.1572079$ & $9.5981991$ & $10.2806717$ & $8.4734818$ & $8.9501754$ & 
$9.6808166$ & $6.6462725$ & $7.5342431$ \\ 
$10$ & $9.7730448$ & $10.2160418$ & $10.8990360$ & $9.0905633$ & $9.5669262$
& $10.2951658$ & $7.2009446$ & $8.0818094$ \\ 
\tableline &  &  &  &  &  &  &  & 
\end{tabular}%
\end{table}

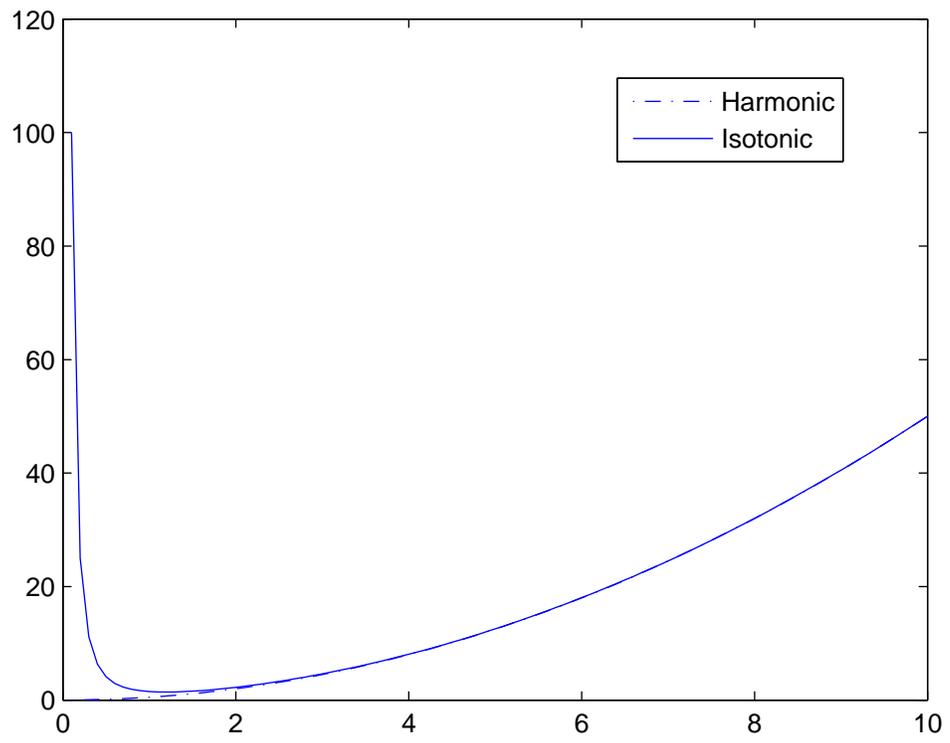
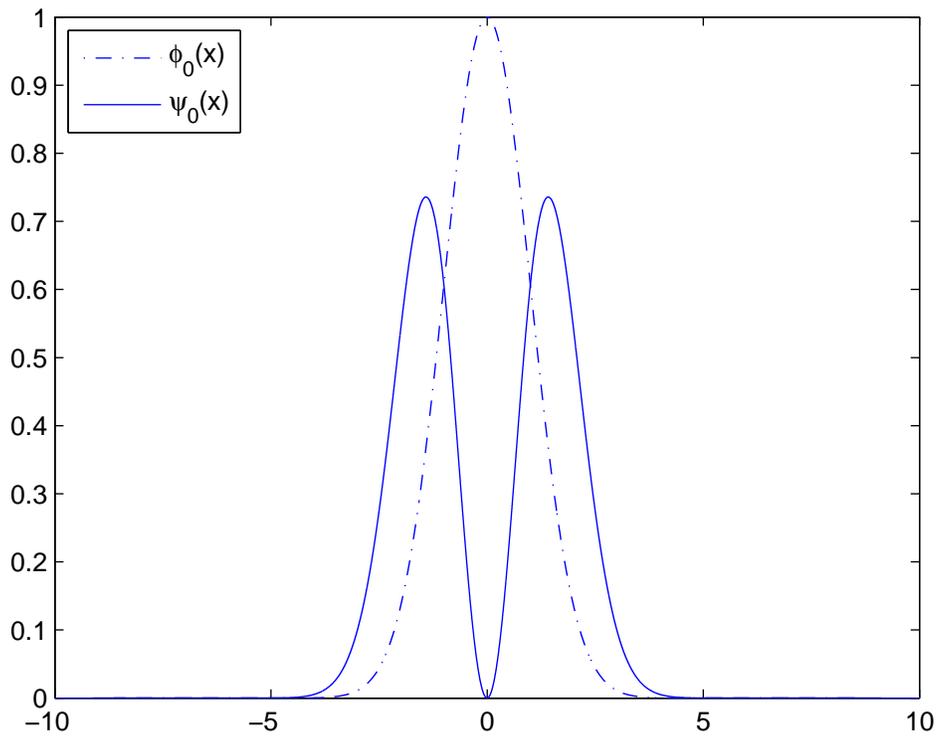
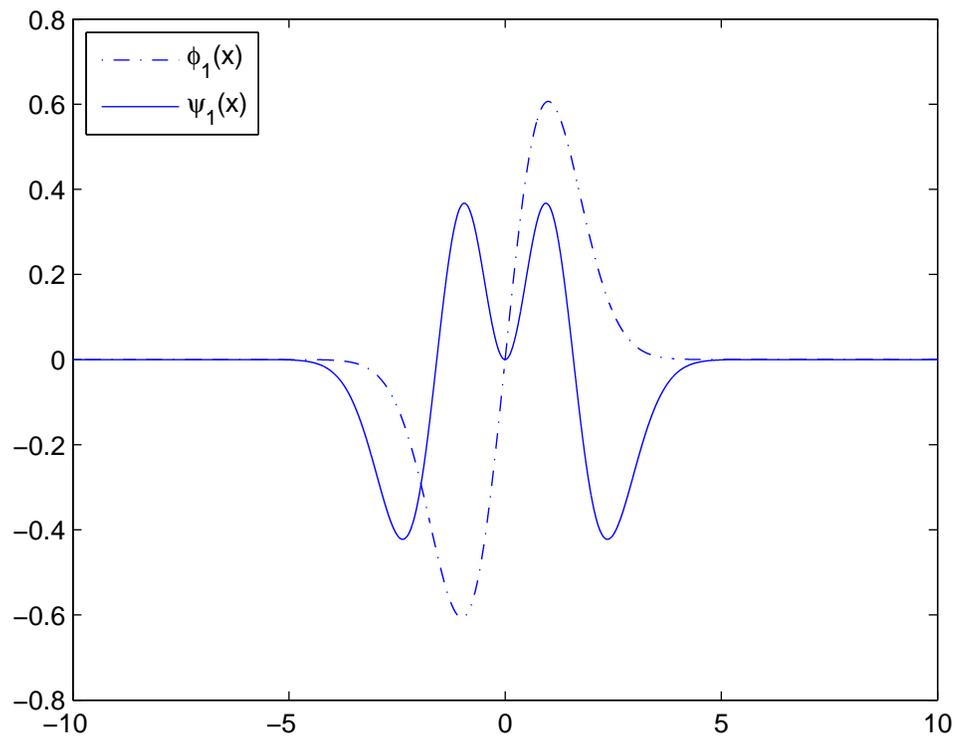
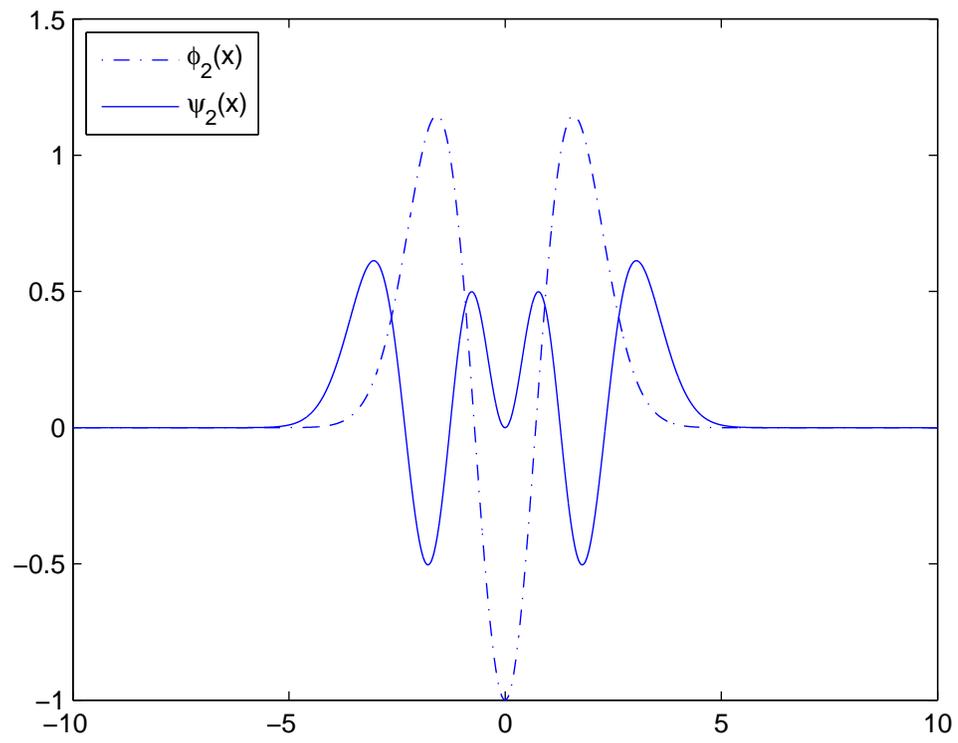
\end{document}